\documentclass[a4paper]{jpconf}
\usepackage{pstricks}
\usepackage{multicol}
\usepackage{array}
\usepackage{graphics}
\usepackage{graphicx}
\usepackage{setspace}
\usepackage{lettrine}
\usepackage{amsmath}
\usepackage{supertabular}

\newcommand{\ket}[1]{\left| #1 \right\rangle}

\newcommand{\sfrac}[2]{\textstyle\frac{#1}{#2}}

\definecolor{dred}{rgb}{0.6,0.0,0.0}

\begin{document}
\title{Weakly-bound rare isotopes with a coupled-channel approach 
that includes resonant levels}

\author{L Canton$^1$, P R Fraser$^2$, J P Svenne$^3$, K Amos$^4$,
S Karataglidis$^5$ and D van der Knijff$^6$}

\address{$^1$ Istituto Nazionale di Fisica Nucleare, 
Sezione di Padova, I-35131, Italy}

\address{$^2$ Instituto de Ciencias Nucleares, Universidad Nacional
Aut\'onoma de M\'exico, 04510 \\ M\'exico D.F., Mexico}

\address{$^3$ Department of Physics and Astronomy, University of
Manitoba, Winnipeg, Manitoba, Canada R3T 2N2}

\address{$^4$ School  of Physics,  University of  Melbourne, Victoria
3010, Australia}

\address{$^5$ Department of Physics, University of Johannesburg,
P.O. Box 524, Auckland Park, 2006, South Africa}

\address{$^6$ Advanced Research Computing, Information Division,
University of Melbourne, Victoria 3010, Australia}

\ead{luciano.canton@pd.infn.it}

\begin{abstract}
The question of how the scattering cross section changes when the spectra 
of the colliding nuclei have low-excitation particle-emitting  resonances 
is explored using a multi-channel algebraic scattering (MCAS) 
method. As a test case, the light-mass nuclear target $^8$Be, 
being particle-unstable, has been considered. Nucleon-nucleus 
scattering cross sections, as well as the spectra of the compound nuclei 
formed, have been determined from calculations that do, and do not, 
consider particle emission widths of the target nuclear states. 
The resonant character of the unstable excited states introduces 
a problem because the low-energy tails of these resonances can intrude 
into the sub-threshold, bound-state region. This unphysical behaviour
needs to be corrected by modifying, in an energy-dependent way, the shape 
of the target resonances from the usual Lorentzian one. 
The resonance function must smoothly reach zero at the elastic threshold. 
Ways of achieving this condition are explored in this paper.
\end{abstract}

\section{Introduction}
\label{Intro}

The advent of radioactive ion beam (RIB) physics prompts the
consideration of new theoretical challenges involving weakly-bound
systems. Radioactive nuclei, especially those close to
the drip lines, can have quite low particle emission thresholds and
consequently have low-lying resonance states in their spectra.
Low-energy scattering of RIB is a coupled-channel
problem that involves such low-lying resonant states 
of the scattered ion.
To address this, a Multi-Channel Algebraic Scattering
(MCAS) formalism~\cite{Am03} is used, in which momentum space solutions 
of coupled Lippmann-Schwinger equations are found,
including also all negative-energy bound states. 
A finite-rank separable representation using
an ``optimal'' set of sturmian functions~\cite{Ra91} serves to 
construct an input matrix of nucleon-nucleus interactions. 
The MCAS method has the ability to locate all compound system resonance 
centroids and widths, regardless of how narrow. 
For full details see reference~\cite{Am03}.
Also, by use of orthogonalizing pseudo-potentials (OPP) in generating 
sturmians, it ensures the Pauli principle is not violated~\cite{Ca05}, 
despite the collective model formulation of nucleon-nucleus interactions used.
Otherwise, some compound nucleus wave functions possess spurious components.

In the following sections, we first briefly summarize the 
MCAS method~\cite{Am03,Ca05}, which has been extended to unstable target states
in Ref.~\cite{Fr08}.
In section II, we display and discuss our results for neutron scattering from 
$^8$Be, with and without the excited target states having non-zero widths.
In section III, we look briefly at the implications of letting the 
target-state widths have energy dependence, to ensure that these 
resonances do not extend to negative energies, where they would produce 
unphysical behaviour for bound, sub-threshold, states. 
Finally, section IV gives our brief concluding remarks.

\section{The MCAS formulation for unstable target states}
\label{MCAS}

$S$-matrix equations in the MCAS methodology take the form:
\begin{equation}
S_{cc'} = \delta_{cc'} - \textrm{i}^{\,l_{c'} - l_c + 1} \pi \mu 
\sum^{N}_{n,n'
= 1} \sqrt{k_c} \; \hat{\chi}_{cn}(k_c) \left( \left[ {\mbox{\boldmath
$\eta$}} - \mathbf{G}_0 \right]^{-1} \right)_{nn'} \hat{\chi}_{c'n'}
\sqrt{k_{c'}}\; .
\label{Smx}
\end{equation}

Traditionally, all target states are taken to have eigenvalues of zero
width. Then the integrals in the (complex) Green's functions are evaluated
using the method of principal parts, where, in the limit 
$\epsilon\rightarrow0$,
the Green's functions take the form
\begin{equation}
\left[ \mathbf{G}_0 \right]_{nn'} = \mu \left[
\sum^{\text{open}}_{c = 1} \int^{\infty}_0 \hat{\chi}_{cn}(x)
\frac{x^2}{ k^2_c - x^2 + \color{red}\textrm{i}\epsilon\color{black} }
\hat{\chi}_{cn'}(x) \, dx \right.
- \left. \sum^{\text{closed}}_{c = 1}
\int^{\infty}_{0} \hat{\chi}_{cn}(x) \frac{ x^2 }{ h^2_c + x^2 }
\hat{\chi}_{cn'}(x) \, dx \right].
\label{S-G}
\end{equation}
\mbox{\boldmath $\eta$}  is a column vector of sturmian eigenvalues
and $\hat \chi$ are form factors determined from the chosen sturmian
functions. 
The factor $\mu$ in Eqs.~\ref{Smx} and \ref{S-G}
is $\mu = 2m_{red} / \hbar^2$, with $m_{red}$ being the reduced mass. 
This method assumes time evolution of target states is given by
\begin{equation}
\ket{x,t} = e^{-\textrm{i} H_0 t/\hbar}\; \ket{x,t_0} =
e^{-\textrm{i} E_0 t/\hbar}\; \ket{x, t_0}.
\end{equation}
However, if states decay, they evolve as

\begin{equation}
\ket{x,t} = e^{-(\Gamma t/2\hbar)}\; e^{-\textrm{i}E_0t/\hbar}\; \ket{x,t_0}.
\end{equation}
Thus, in the Green's function, channel energies become complex,
as do the squared channel wave numbers,
\begin{equation}
\hat{k_c}^2 = \mu \left(E - \epsilon_c + \textrm{i}\sfrac{\Gamma_c}{2}
\right) \: ; \:\:
\hat{h_c}^2 = \mu \left( \epsilon_c - E - \textrm{i}\sfrac{\Gamma_c}{2}
\right) ,
\label{hatkandhath}
\end{equation}
where $\sfrac{ \Gamma_c }{2}$ is half the width of the target state
associated with channel $c$. The Green's function matrix
elements are then
\begin{equation}
\left[ \mathbf{G}_0 \right]_{nn'}  = \mu \left[
\sum^{\text{open}}_{c = 1} \int^{\infty}_0 \hat{\chi}_{cn}(x)
\frac{x^2}{ k_c^2 - x^2 + \color{red}\textrm{i} \mu \sfrac{\Gamma_c }{2}
\color{black} } \; \hat{\chi}_{cn'}(x) \, dx  \right.
- \left. \sum^{\text{closed}}_{c = 1}
\int^{\infty}_{0} \hat{\chi}_{cn}(x)
\frac{ x^2 }{ h_c^2 + x^2 - \color{red}\textrm{i} \mu \sfrac{\Gamma_c }{2}
\color{black}} \; \hat{\chi}_{cn'}(x) \, dx \right] .
\label{S-Ggamma2}
\end{equation}
When poles are moved significantly off the real momentum axis,
integration along this axis is feasible.

\section{The case of $^9$Be} 
\label{Be9}
The low excitation spectrum of $^8$Be has a $0^+$ ground
state that decays into two $\alpha$-particles
(width: $6 \times 10^{-6}$ MeV), a $2^+$ resonance state (centroid:
3.03	MeV; width: 1.5 MeV) and a $4^+$ resonance state (centroid: 11.35 MeV; width: ${\sim}3.5$ MeV). In Table~\ref{Be8-param} are given the parameters
in the channel-coupling potentials, as well as the $\lambda^{OPP}$ 
values giving the strengths of Pauli hindrance.
\begin{table} [ht] \centering
\caption
{\label{Be8-param} 
The parameter values defining the nucleon-$^8$Be interaction.}
\begin{supertabular}{>{\centering}p{28mm} p{27mm}<{\centering} 
p{27mm}<{\centering}}
\hline
\hline
 & Odd parity & Even parity\\
\hline
$V_{\rm central}$ (MeV) & -31.5 & -42.2 \\
$V_{l l}$ (MeV)       & 2.0   & 0.0 \\
$V_{l s}$ (MeV)       & 12.0 & 11.0 \\
$V_{ss}$ (MeV)        & -2.0 & 0.0 \\
\end{supertabular}
\begin{supertabular}{>{\centering}p{19.5mm} p{21mm}<{\centering} 
>{\centering}p{21mm} p{19mm}<{\centering} }
\hline
\hline
Geometry & $R_0 = 2.7$ fm & $a = 0.65$ fm & $\beta_2 = 0.7$ \\ 
\end{supertabular}
\begin{supertabular}{>{\centering}p{28mm} p{27mm}<{\centering} 
p{27mm}<{\centering}}
\hline
\hline
                        &$0s_{1/2}$ &$0p_{3/2}$\\
\hline
 $0^+ \; \lambda^{(OPP)}$ & 1000 & 0.0 \\
 $2^+ \; \lambda^{(OPP)}$ & 2.0 & 0.2 \\
 $4^+ \; \lambda^{(OPP)}$ & 0.0 & 0.0 \\
\hline
\hline
\end{supertabular}
\end{table}
Two evaluations of the $n + ^8$Be cross section are obtained using
this spectrum; one with target-state widths set to zero, and the
other taking into account the widths of the excited levels.
The two results will be identified by the terms `no-width'
and `width', respectively.
Both cases are calculated with the same nuclear interaction,
taken from a rotor model~\cite{Fr10}. 
Table~\ref{Tab2} displays $^9$Be spectrum results from these calculations. 

\begin{table}[ht]
\begin{center}
\caption{\label{Tab2}
Widths of resonances in $n+^8Be$ scattering for the no-width and 
width cases, compared to experimental values. 
Boldface entries highlight the matches we find between theory and
experiment to within a factor of 3.}
\setstretch{1.35}
\normalsize
\begin{tabular}{cc|cc|cc}
\hline
\hline
$J^\pi$ & $\Gamma_{exp.}$ & $\Gamma_{no-width}$
& $\sfrac{\Gamma_{no-width}}{\Gamma_{exp.}}$
& $\Gamma_{width}$
& $\sfrac{\Gamma_{width}}{\Gamma_{exp.}}$\\
\hline
$\sfrac{1}{2}^+$ & 0.217$\pm$0.001     & --- & ---  & 1.595 & 7.350\\
$\sfrac{7}{2}^-$ & 1.210$\pm$0.230     & 2.08$\times10^{-5}$
& 1.72$\times10^{-5}$ & 1.641 & \color{dred}\textbf{1.356}\\
$\sfrac{1}{2}^-$ & 1.080$\pm$0.110     & 0.495 & 
\color{dred}\textbf{0.458}\color{black}
& 1.686 & \color{dred}\textbf{1.561}\\
$\sfrac{5}{2}^+$ & 0.282$\pm$0.011     & 0.187 & 
\color{dred}\textbf{0.663}\color{black}
& 0.740 & \color{dred}\textbf{2.624}\\
$\sfrac{3}{2}^-$ & 1.330$\pm$0.360     & 0.466 & 
\color{dred}\textbf{0.350}\color{black}
& 3.109 & \color{dred}\textbf{2.337}\\
$\sfrac{5}{2}^-$ & $7.8{\times}10^{-4}$ & 0.060 & 76.74
& 2.772 & 3554\\
$\sfrac{9}{2}^+$ & 1.330$\pm$0.090     & 0.386 & 0.290
& 2.498 & \color{dred}\textbf{1.878}\\
$\sfrac{3}{2}^+$ & 0.743$\pm$0.055     & 3.286 & 4.423
& 5.162 & 6.947\\
\hline
\hline
\end{tabular}
\setstretch{1.0}
\end{center}
\end{table}
\begin{figure}[ht]
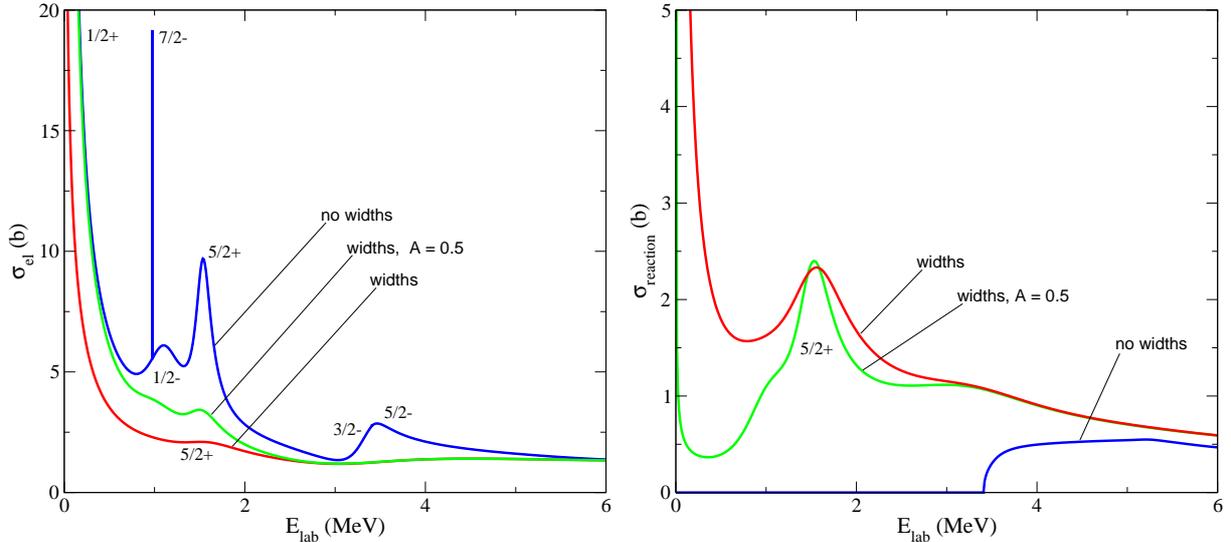

\centering
\resizebox{1.0\columnwidth}{!}{%
\includegraphics*{Fig1-elast.eps} \ \ \ \  
\includegraphics*{Fig2-react.eps}}
\caption{Elastic (left panel) and reaction (right panel) 
cross section as a function of neutron laboratory energy
with three treatments of the widths of target states (see text).}
\label{fig1}
\end{figure}

Taking the excited states of $^8$Be to be resonances gives the same spectral 
list as when they are treated as zero-width, but the evaluated widths of 
the compound nuclear states significantly increase, as reflected in the 
cross sections. These increases bring the theoretical $^9$Be state widths 
closer, often significantly, to experimental values, as evidenced by 
comparison of the final columns of the two tables. 
Only the $\frac{5}{2}^-$, also poorly recreated in centroid, 
has a worse match with widths applied.
The resultant elastic and reaction cross sections are shown in
figure \ref{fig1}. The blue line depicts the results 
without widths, the red line with constant widths multiplied by the 
Heaviside (step) function: 
$H(E) = 0$ for $E \le 0.0$; $H(E) = 1$ for $E > 0.$
(For the green line, see section~\ref{scaling}.) 


Introducing target state widths, the resonances in $^9$Be are
suppressed but still present; their widths increasing and magnitudes
decreasing so as all but the $\left(\sfrac{5}{2}\right)^+_1$, and arguably
$\left(\sfrac{5}{2}\right)^+_2$, cannot be discerned from the background.
The reaction cross section found in the no-width case is zero until
3.4	MeV (3.03 MeV in the centre of mass frame), the first inelastic threshold.  When the target state has width different from zero the elastic cross section 
has flux loss from zero projectile energy upwards. There is an asymptotic 
behaviour as energy approaches zero.  In the reaction cross section from 
the width calculation, there is a broad peak at ~1.5 MeV, corresponding 
to the ${\textstyle\frac{5}{2}}^+$ resonance. Using widths not
varying with energy (see the next section), the bound states, $E < 0.0$,
of the compound $^9$Be system are unstable due to 
the couplings to the decaying excited states of the $^8$Be core.  
For this reason the widths have been multiplied by a Heaviside 
function to ensure that the compound system is stable at negative energies. 
An improvement of this technique is discussed in the next section.

\section{Energy dependence of widths}
\label{scaling}

A smooth description of the energy dependence of the
decay width is needed to avoid any pathological behaviour caused by
the Heaviside function.
This function allows maintaining the stability of the compound system
in the sub-threshold region, by setting to zero all widths for $E < 0$,
but can generate ``ghost'' levels where the application of widths
could move a sub-threshold state above $E=0$, thus creating two copies
of the same state. Also, a dramatic overestimation
of the reaction cross-section occurs in the region close to threshold.

We therefore need to improve the model with a suitable parametrization
of the energy-dependence of the decaying width for the excited levels,
with the constraint that the width must be zero at the scattering threshold,
and  equal to the experimental width at the resonance centroid energy.
We are presently exploring various forms of energy dependences for the 
widths of these excited states ($\Gamma_c(E)=\Gamma_c \times U(E)$).
Examples of such expressions are
\begin{equation}
U(E) = \frac{(1+A)}{(1+A)^2-1}\left\{ \frac{1+A}{A(1-x)^2+1}- 
\frac{1}{A(x^2)+1}\right\} H(E)
\label{rational}
\end{equation}
where $x= E/E_r$, $H(E)$ is the Heaviside function defined above
and $A$ is a parameter, and the Gaussian form
\begin{equation}
U(E) = e \left(\frac{E}{E_r}\right)^2 e^{-\left(\frac{E}{E_r}\right)} H(E).
\end{equation}
In this paper we study only the first form. The second, and others, 
will be explored in future work. 
Also yet to be taken into account is causality, 
as needed when dealing with energy-dependent widths~\cite{Ma86}.

\begin{figure}[ht]
\centering
\resizebox{0.70\columnwidth}{!}{%
\includegraphics*{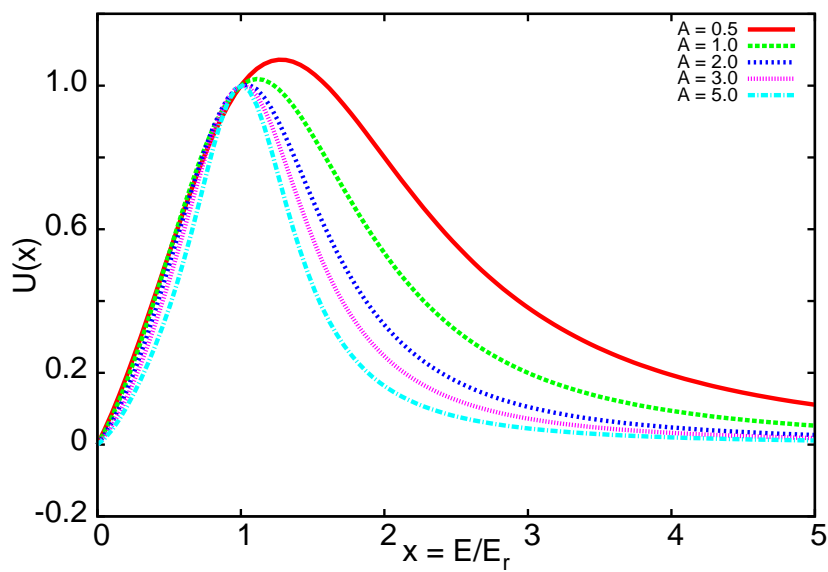}}
\caption{The scaling function, $U(E/E_r)$, 
for various values of the parameter A.}
\label{fig3}
\end{figure}

In figure~\ref{fig3} we give the induced energy dependence for 
the function $U(x=E/E_r)$ with respect to the renormalized energy $x$. 
The curves differ for different values of the parameter $A$.
The green line in figure \ref{fig1} shows
the effect of this energy-dependent width on elastic and 
reaction cross sections, respectively, when $A=0.5$.
Clearly, for this value of $A$, the asymptotic behaviour in the
reaction cross section at $E = 0.0$ is suppressed,
but not resolved.
The use of an energy dependence in the widths of the excited states of the 
target introduces a new aspect that has to be taken into consideration. 
In principle, an energy dependence in the widths induces a corresponding 
energy-dependence in the positioning of the resonance centroids. This 
additional effect is a direct consequence of the principle of causality 
which demands that the real and imaginary part of any Green's function 
matrix elements are related by a dispersion integral. Such an aspect has been 
extensively discussed in the study of the properties of the nuclear optical 
potential~\cite{Ma86}. Eq.~\ref{rational} is particularly 
convenient in this regard, because the corresponding dispersion integral 
leads to an analytical energy-dependent shift of the resonance centroids.
Such analytical forms have been derived in Ref.~\cite{Ca91} for a class 
of functions  which includes Eq.~\ref{rational}. The resulting energy
shift function $\Delta S_c(E)=\Gamma_c \times S(E)$ is given by the analytical 
expression:
\begin{equation}
S(E)= \frac{1}{A(x-1)^2+1}\left\{ c_1-(x-1)c_2\right\} +
\frac{1}{A(x)^2+1}\left\{ c_3+(x)c_4\right\}
-\frac{\ln|x|}{\pi} U(x)
\end{equation}
where
\begin{equation}
c_1 = \frac{(1+A)^2}{(1+A)^2-1} \frac{\ln\sqrt{1+\sfrac{1}{A}}}{\pi}
 \: \: \:; \: \: \:
c_2 = \frac{(1+A)^2}{(1+A)^2-1} \frac{\sqrt{A}}{\pi} 
\left(\frac{\pi}{2}+arctg{\sqrt{A}}\right) \:,
\end{equation}
and 
\begin{equation}
c_3 = \frac{(1+A)}{(1+A)^2-1} \frac{\ln\sqrt{A}}{\pi}
 \: \: \:; \: \: \:
c_4 = \frac{(1+A)}{(1+A)^2-1} \frac{\sqrt{A}}{2} \: .
\end{equation}

\section{Conclusions}
\label{concl}

An extension to the multi-channel algebraic scattering (MCAS) 
formalism that considers particle-decay widths of target nucleus 
eigenstates has been applied to a range of light mass nuclear 
targets~\cite{Fr10}, in addition to the results for $n + ^8$Be 
shown in this paper.
To ensure that any resonance aspect of target states does not
have influence below the nucleon-nucleus threshold in our formalism,
we have used a Heaviside function to cut off tails of the
positive-energy resonances below zero energy of the compound system.
This procedure, however, is too simplistic since it introduces
non-physical singularities in cross sections at the approach
to zero energy, and may also incorrectly overestimate reaction cross
just above the threshold.

A better energy dependent scaling factor is needed:
one with value one at the centroid energy and approaching zero
for lower and higher energies.
We propose here possible means of achieving such modified Lorentzians,
and show, for one example, their effect on the elastic and reaction 
cross sections in the scattering of neutrons from $^8$Be. 
Work is in progress to apply these and other forms to the 
light-nuclear systems considered in our previous work.
Work is in progress, also, for preserving causality in the presence
of energy-dependent widths by the use of appropriate dispersion relations.

\ack 
This work is supported by the National Research Foundation, South Africa, 
the Natural Sciences and Engineering Research, Council, Canada, 
INFN, Italy and the Australian Academy of Science.

\end{document}